\documentclass[conference]{IEEEtran}
\IEEEoverridecommandlockouts
\usepackage{cite}
\usepackage{url}

\usepackage{booktabs}
\usepackage{algorithmic}
\usepackage{graphicx}
\usepackage{textcomp}
\usepackage{xcolor}
\usepackage{gensymb}
\usepackage{pgfplots}
\usepackage{tikz}
\usetikzlibrary{plotmarks}
\usepackage{soul}  
\usepackage[normalem]{ulem}
\pgfplotsset{compat=1.11}
\newlength\fwidth

\usepackage{calligra}
\usepackage{mathtools,xparse}
\usepackage{eucal}
\usepackage{amsmath,amsfonts,amssymb,amsthm,amssymb}
\usepackage{bbm}
\usepackage{commath}
\usepackage[mode=buildnew]{standalone}

\begin{document}

\title{Representation Learning Strategies to Model Pathological Speech: Effect of Multiple Spectral Resolutions}

\author{Gabriel Figueiredo Miller$^1$, Juan Camilo Vásquez-Correa$^{1,2}$, Juan Rafael Orozco-Arroyave$^{1,2}$, Elmar N\"oth$^1$\\
$^1$\textit{Pattern Recognition Lab. Friedrich-Alexander-Universit\"at Erlangen-N\"urnberg, Germany}\\
$^2$\textit{GITA Lab. Faculty of Engineering, University of Antioquia UdeA, Medellín, Colombia}\\
corresponding: \url{gabriel.f.miller@fau.de}, \url{juan.vasquez@fau.de}
}

\maketitle

\begin{abstract}
This paper considers a representation learning strategy to model speech signals from patients with Parkinson's disease and cleft lip and palate. In particular, it compares different parametrized representation types such as wideband and narrowband spectrograms, and wavelet-based scalograms, with the goal of quantifying the representation capacity of each. Methods for quantification include the ability of the proposed model to classify different pathologies and the associated disease severity. Additionally, this paper proposes a novel fusion strategy called multi-spectral fusion that combines wideband and narrowband spectral resolutions using a representation learning strategy based on autoencoders. The proposed models are able to classify the speech from Parkinson's disease patients with accuracy up to 95\%. The proposed models were also able to asses the dysarthria severity of Parkinson's disease patients with a Spearman correlation up to 0.75. These results outperform those observed in literature where the same problem was addressed with the same corpus. 

\end{abstract}


\section{Introduction}

Automated disease prognosis and classification has been extensively explored by the research community for its potential benefits in tele-medicine with respect to cost, time efficiency and public availability~\cite{HZ_PDVocal,MAL_suitDysphoniaMeasurementsTelemonitoringPD,JG_telemedInternationalPerspective}. These benefits have been observed for many diseases, specifically ones where bio-markers can be identified and monitored effortlessly and consistently, relative to visiting a healthcare provider~\cite{AE_dermatologistClassSkinCancer,YNZ_canSmartphoneDiagnosePD,MG_autoIDHypernasality}. Additionally, automation of general health care prognostics has allowed for a more objective and consistent approach to treating patients, putting to use volumes of data and bringing to light insights that have pushed the boundaries of what was thought possible with modern medicine~\cite{NGM_machineLearningImprovingDiagnosis}. 

One interesting sub-field of diagnostic scenarios where this has played out is the assessment of diseases that affect the speech production system. It is known that speech signals contain paralinguistic information containing speciﬁc cues about a given speaker, e.g., the presence of diseases that may alter their communication ability~\cite{BS_compParalinguistics,BS_affectiveBehaveCompute,NC_speechAnalysisForHealth}. Generally, researchers look at clinical observations in the speech of patients and try to objectively and automatically measure two main aspects of a given disease: (1) the presence of a disease via classification of healthy control (HC) subjects and patients and (2) the level of degradation of the speech of patients according to a specific clinical scale~\cite{JROA_charMethodsDetectingMultipleDisorders}. 

Generally, unsupervised methods based on feature representation learning have the potential to extract more abstract and robust features than those manually computed. These features help to improve the accuracy of different models to classify pathological speech. There are many recent studies focused on extracting features based on deep learning strategies for assessment of pathological speech. In~\cite{TG_assessingDegreeNativenessParkinsonsCondition}, the authors detail a model used to win the "2015 computational paralinguistic challenge (ComParE)"~\cite{BS_compParalinguistics} where the challenge was to build an automatic estimator of the neurological state of Parkinson's disease (PD) patients. The winning model was able to predict the neurological state of a PD patients with a 0.65 Spearman correlation ($\rho$) coefficient. The model used Gaussian processes and deep neural networks (DNNs) to perform the prediction of the clinical score. In~\cite{YNZ_canSmartphoneDiagnosePD}, phonation and articulation-based features were distilled into bottleneck features using autoencoders (AEs), which provide an unsupervised learning framework to infer an optimal representation of a set of features~\cite{LD_recentAdvancesDLSpeechResearch}. The bottleneck features were then fed into a K-nearest neighbor (KNN) classifier~\cite{ZZ_introMLKNN}. The authors reported a classification accuracy of 94\%, though the hyper-parameters of the AE were optimized on the test set of PD speakers, which make the results optimistic.  Another approach that uses a parallel representation learning strategy to model speech signals from patients with different speech disorders (including cleft lip and palate (CLP) and PD) was developed in~\cite{JCVC_parallelRepresentationLearning}. The paper tested convolutional and recurrent AE frameworks in order to obtain a hidden representation in the bottleneck space which was then used to classify the presence of the speech disorders. The model also utilized the reconstruction error of the AE in different spectral components of the speech signal as part of the feature set. The proposed models were accurate in modeling speech signals from patients with both PD and CLP. In particular, the authors reported accuracy up to 84\% in the classification of PD vs. HC speakers and 97\% for CLP vs. HC.

This paper goes beyond the work done in~\cite{JCVC_parallelRepresentationLearning}, and considers the effect that different parametrized inputs have in modeling and analyzing speech disorders that specifically affect PD. Three parametrized speech representations are considered: wideband and narrowband spectrograms, and wavelet-based scalograms (note that in ~\cite{JCVC_parallelRepresentationLearning}, no signal parameter specifications were given, e.g., the analysis window of the spectrograms used as input to the AEs). Additionally in this paper, fusion strategies are presented. With these strategies it is shown that the different spectral representations are complementary to each other in modelling the speech disorders of PD speakers. Of the fusion strategies presented, we propose a novel fusion-based approach, which we cal the multi-spectral autoencoder. It combines information that is better captured by each speech representation (e.g., wideband, narrowband or wavelet representations) by first distilling the feature set produced by each representation separately using several convolutional layers. These distilled feature sets are then concatenated and convoluted a final time to produce a feature set that contains qualities of PD speech that each representation uniquely captures. This approach is shown to be superior in it's classification ability to traditional approaches tested in this study, and found in similar studies. 


\section{Methods}

\subsection{Speech Representations} \label{speech_rep_types}

\begin{figure}[!ht]
    \centering
    \includegraphics[width=\linewidth]{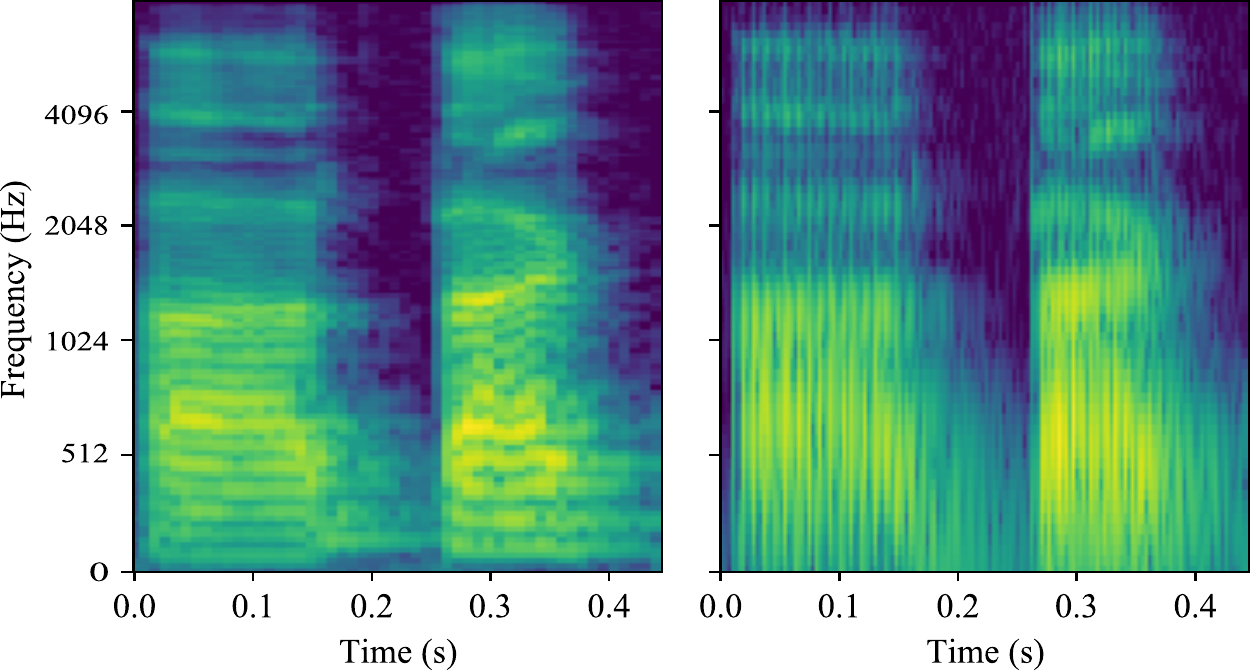}
    \caption{The spectrogram shows the production of the Narrowband (left) and wideband (right) signals, characterized by horizontal and vertical striations respectively. Both are computed for 500\,ms length speech segments and used as input to train the CAEs.}
    \label{fig:spectrograms}
\end{figure}

Three different representation types were considered in this analysis: wideband and narrowband spectrograms and wavelet-based scalograms. Each representation is known to emphasize different parts of the speech signal. Considering first the wideband and narrowband signals, each one helps to optimize between the time and frequency resolution of the spectrogram representation. 
The wideband spectrogram is valued for its quick temporal response and is generally used for word-boundary location and formant tracking, giving more emphasis to obtain articulatory information from the signal. On the other hand, the narrowband spectrogram is preferred to measure the fundamental frequency due to the fact that it manifests typically as horizontal striations in the spectrogram, thus giving more emphasis to extract prosodic cues from the spectrogram~\cite{EM_timeDomainFreqDomainTechsSpeechMod,GF_acousticTheorySpeechProduction}. The narrowband spectrogram is obtained by computing the short time Fourier transform (STFT) using a window size of 30\,ms, a time shift of 10\,ms and 128 Mel frequency bands. Similarly, in order to derive the wideband spectrogram, the STFT was computed using a window size of 5\,ms, a time shift of 3\,ms and 64 Mel-frequency bands. The difference between both spectrograms can be observed in Figure~\ref{fig:spectrograms}. Note that a bi-cubic interpolation is applied on both wideband and narrowband representations to ensure each frame is of uniform length ($128\times126)$ \cite{GB_smoothSurfaceInterpolation}.

The wavelet transformation is an alternative speech representation that provides both time and frequency information with varying resolutions at different decomposition levels~\cite{SM_wvltTourSP,DWR_waveNetsScaleEquivariantLearning}. It can be beneficial relative to the short-time Fourier transform particularly when there are dramatic changes in the signal being analyzed. Specifically, the continuous wavelet transform utilizes a Gaussian function modulated by a complex exponential to analyze the frequency content of a signal~\cite{YNJ_newSpectrogramEvaluatedEnhancedCWT}. Properties of the Gaussian function are particularly useful when it comes to minimizing the spread in the time-frequency plane as the Gaussian window yields the smallest possible root mean square frequency variance for a given temporal resolution.

\subsection{AE-Based Features}

AEs are a type of neural network designed to extract features from unlabelled data. Hence, they are able to detect and remove input redundancies in order to preserve only essential aspects of the data into robust and discriminative representations~\cite{JM_stackedCAEsHierarchicalFeatureExtraction}.
AEs have been cited as a robust method for speech analysis relative to supervised methods, particularly in cases where the specific speech data is not widely available such as in the case of pathological speech modeling~\cite{DJ_semisupervisedAEsSpeechEmotionRecognition,SA_s2sAEsUnsupervisedRepresentationLearning,MT_selfSupervisedAudioRepresentationLearning}. Additionally with supervised methods, it takes a considerable effort to extract traditional speech features, which can be very specific to the problem they are designed for. This is in contrast to AEs and more generally speaking, unsupervised representation learning methods, which learn general purpose speech representations that can be adapted to different problems~\cite{MT_selfSupervisedAudioRepresentationLearning}.

Different speech features can be extracted from trained AEs. Below two such sets that were used are described.

\subsubsection{Bottleneck Representation} 

The bottleneck representations have shown to be effective in modelling speech. Specifically, bottleneck features benefit from the non-linear compression feature of deep learning architectures and also maintain the specific benefit that a given learning architecture offers~\cite{TS_aeBottleneckFeaturesUsingDBN}.

\subsubsection{Reconstruction Error} 

The main hypothesis about considering these features is that not all frequency regions of the spectral representation can be reconstructed with the same error, and such a reconstruction error is related to the presence of paralinguistic aspects such as different speech disorders. This is motivated from an anomaly detection paradigm, since the AEs are trained with information from healthy speech, thus it is expected that the reconstruction error of pathological speech signals would be different than the observed for healthy speech~\cite{CF_autoencoderUnsupervisedDataAnalyticsAnomalyDetection,DP_multimodalAnomalyDetectorRobots,JCVC_parallelRepresentationLearning}. For instance, in the case of PD patients it is well known that dysarthria causes monotonicity and monoloudness, which means that the corresponding signal has less variability than observed for healthy people. Thus, it is easier to reconstruct PD speech by the AEs than healthy speech~\cite{JCVC_parallelRepresentationLearning}. Conversely, for CLP patients that have problems producing phonemes with more energy content in higher frequencies, like sibilants, their speech can be more difficult to reconstruct than speech from typical speakers by AEs. 

\subsection{Fusion-Based Frameworks}

The bottleneck and reconstruction error features are derived for each speech representation: narrowband and wideband spectrograms and the wavelet scalogram, and then used to build a classifier for discerning PD or CLP patients from HC speakers. In addition, features associated with each representation are combined together to leverage each individual representation's benefits. Both early and late fusion strategies are considered to combine the different speech representations. In the early fusion strategy, the features obtained for each speech representation are concatenated and used in a later classification algorithm. The late fusion approach consists of  building separate classifiers using each speech representation and combining the scores assigned by each classifier into a global decision. Individual classifiers are based either on a neural network or a support vector machine (SVM). The fusion is made based on a weighted sum of the scores of each classifier in the training set. The weights for the fusion are found using a stochastic gradient descent classifier with a Hinge loss function and an $L_2$ regularization strategy. Higher weights are given to the representation that was better at classifying the training set~\cite{JT_sgdLargeScaleSVM,AK_simpleWeightDecayImproveGeneralization}. 

\begin{figure*}
    \centering
    \includegraphics[scale=.9]{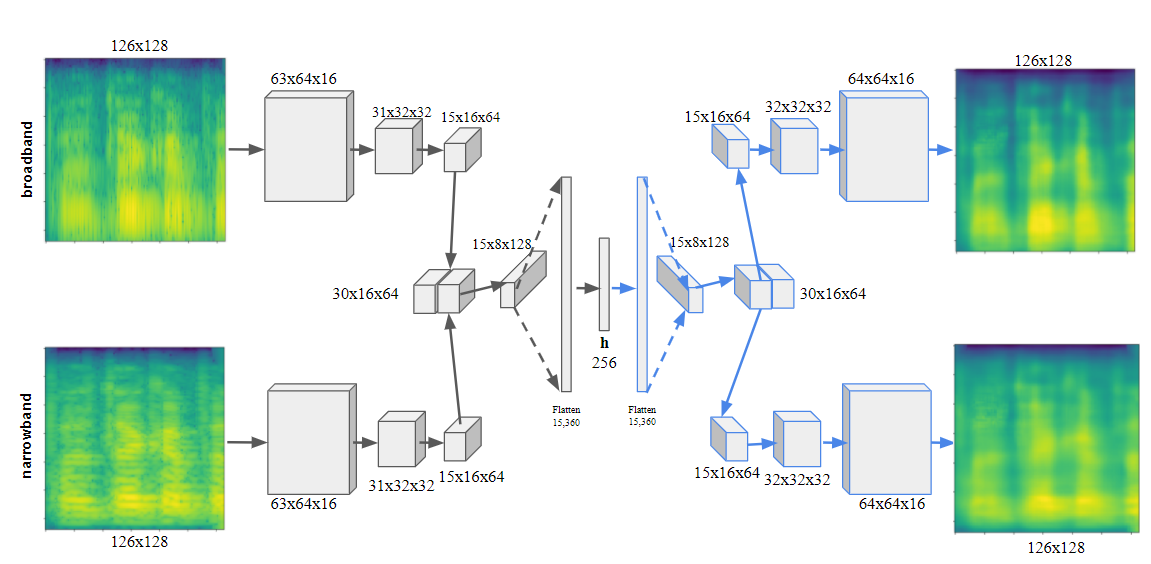}
    \caption{Multi-spectral convolutional autoencoder scheme. \textbf{FC}: fully connected layer, \textbf{h}: bottleneck representation.}
    \label{mcCAE_architecture}
\end{figure*}

A third approach to combine the speech representations is proposed in this study and called multi-spectral fusion. The approach takes advantage of a convolutional AE which distills information from multiple signals in parallel and yields a joint bottleneck representation that holds key information from each speech representation (see Figure \ref{mcCAE_architecture}). To obtain the feature set using the multi-spectral AE architecture, a pair (wideband and narrowband) or triplet (wideband, narrowband, and wavelet) of signals are mapped to lower dimensional spaces in parallel. After this is done three times, i.e., each representation passes through three convolutional layers, the tensors associated with each representation are then concatenated together. The bottleneck features are then derived by passing this set of features through the fourth and final layer. The original signals are similarly reconstructed, namely the bottleneck features are up-sampled, evenly split and then up-sampled until the original two or three representations are obtained. Reconstruction errors are obtained for each representation.

\section{Data Description}


\subsection{TED-X Spanish Corpus}

The TED-X Spanish Corpus (TSC) contains over 24 hours of speech obtained from TED-X talks (11,243 audio files in total)~\cite{CDHM_tedxSpanishCorpus}. This was the corpus used to train the AEs. The speakers are all native Spanish speakers and for the most part are men (102 males, 40 females). Due to the fact that the set of PD and CLP speakers are more evenly distributed, the gender bias in the TSC is addressed by evenly sampling from the male and female speaker pools. Each speaker in the dataset has several clips from talks with a duration between three and ten seconds approximately.

\subsection{PC-GITA Corpus}
The considered models are evaluated on the PC-GITA corpus~\cite{orozco2014new}. The data comprises of utterances from 50 PD patients and 50 HC subjects, Colombian Spanish native speakers. The participants were asked to pronounce 10 sentences, six diadochokinetic (DDK) exercises, one text with 36 words, the sustained phonation of vowels, and a monologue. For this study, only the DDK utterances were considered. The speech recordings were collected in a soundproof-booth with a professional directed microphone.
All patients were evaluated by a neurologist expert according to the MDS-UPDRS-III scale~\cite{goetz2008movement}, and they were recorded in ON state. The dysarthria severity of the participants was evaluated according to the m-FDA scale~\cite{vasquez2018towards}, which consists of 13 items and evaluates seven aspects of the speech including breathing, lips movement, palate/velum movement, laryngeal movement, intelligibility, and monotonicity. Each item ranges from 0 to 4 (integer values), thus the total score ranges from 0 (healthy speech) to 52 (completely dysarthric). Table~\ref{tab:metadata} shows clinical and demographic information from the PD patients and HC subjects that participated in this study.

\begin{table}[!ht]
\centering
\caption{Demographic information from the PD patients that participated in this study.}
\label{tab:metadata}
\resizebox{.9\linewidth}{!}{
\begin{tabular}{llll}

\toprule
\textbf{}                                 & \textbf{PD (n=50)} & \textbf{HC  (n=50)} & \textbf{PD vs. HC}   \\
\midrule
Sex (F/M)                      & 25/25              & 25/25               & –           \\
Age                              & 61.0 (9.3)         & 61.0 (9.4)          & 0.49$^a$     \\
Years since diagnosis            & 10.6 (9.1)         & –                   & –             \\
MDS-UPDRS-III                    & 37.7 (18.1)        & –                   & –           \\
MDS-UPDRS-speech                 & 1.3 (0.8)          & –                   & –             \\
m-FDA                     & 28.8 (8.3)         & 8.5 (7.4)           & $\ll$0.005$^a$  \\
\bottomrule
\multicolumn{4}{l}{$^a$p-values calculated using Mann-Whitney U tests}
\end{tabular}}
\end{table}

\subsection{PD Independent Data Set}

A second corpus from PD patients is included for an independent analysis of the multi-spectral AEs to validate the generalization capabilities of the considered approaches, specifically the multi-spectral AEs. The dataset contains 20 PD patients (11 female) and 20 HC speakers (9 female). Again all speakers were Colombian and performed the same exercises as the speakers of the PC-GITA. The mean and standard deviation age for female and male PD speakers was 57.9 (16.9) and 65.3 (7.48), respectively. For the HC speakers, the age distribution for female and male speakers was 60.7 (16.9) and 64.2 (9.8), respectively. The speech material was collected with a noise cancelling headset. The acoustic conditions in this corpus are different than the ones in the PC-GITA corpus in order to evaluate the considered models in real environments that can be found in a normal medical room. 

\subsection{CLP Data}\label{clp_data}

Data for CLP was provided by \emph{Grupo de procesamiento y reconocimiento de se\~nales (GPRS)} of the National University from Colombia. The data contain utterances from 118 children with repaired CLP and 58 HC. The age of the children ranges from 5 to 15 years old. 
All the patients were evaluated by speech therapists and they were diagnosed with hypernasal speech, mainly because changed nasality is often still present after surgical therapy of the cleft. 
The tasks performed by the participants include the pronunciation of isolated Spanish words such as /bola/, /chuzo/, /coco/, /gato/, /jugo/, /mano/, /papa/, and /susi/. These words contain different groups of phonemes to characterize properly the different place and manner of articulation of the patients~\cite{orozco2016automatic}.

\section{Procedural Overview}

To train the autoencoders, audio files from the TSC were split into a training and validation set~\cite{CDHM_tedxSpanishCorpus}. As noted, within the dataset, speakers generally have multiple clips from different segments of their talk. To maintain speaker independence between training and validation sets, segments from the same speaker were grouped together.

The different speech representations used to train the AEs were obtained for segments with 500 ms length. The Phonet toolkit\footnote{https://github.com/jcvasquezc/phonet} was used to predict the posterior probability of each speech file to be plosive (i.e., /p/, /t/, or /k/)~\cite{JCVC_phonet} (probabilities predicted every 25 ms). This precaution was taken to ensure that the segments are always aligned with the position of the plosive sounds in the DDK utterances of the PD corpus. For the CLP data, only a subset of data was used, namely only single utterances of the Spanish words detailed in Section \ref{clp_data}. Thus alignment was not needed. In both cases, the 500 ms speech segments were used to train the AEs.

Extracted features from the AEs were then used to train two classifiers: (1) a fully connected two-layer deep neural network (DNN), and (2) an SVM classifier with a Gaussian kernel. The performance of the considered models was evaluated using a nested 10-fold cross-validation strategy with an internal 9 folds for hyper-parameter optimization based on the accuracy of the training set. Specifically, in the case of PD patients an additional test is performed by training the best performing model with the PC-GITA corpus and using the additional PD corpora as an independent blind test set.

\section{Results}

\subsection{Parkinson's Disease}

Table~\ref{pd_results_table1} shows the results classifying PD patients and HC subjects from the PC-GITA corpus using the different considered speech representations and classifiers. The table also includes the Spearman correlation ($\rho$) between the scores of the classifiers e.g., the posterior probability of the DNN and the distance to the hyperplane of the SVM with respect to the dysarthria severity of the patients according to the m-FDA scale~\cite{vasquez2018towards}.

\begin{table}[ht!]

    \caption{Classification results for PD patients in the PC-GITA corpus. Early and late fusion results for the SVM classifier are reported for the combination of wideband and narrowband features (x2) and (x3) implies the inclusion of the wavelet-based features.}
    \centering
    \scriptsize
    \resizebox{\linewidth}{!}{
    \begin{tabular}{lccccc}
    \hline
         & \textbf{Accuracy}  & $\rho$ & \textbf{AUC} & \textbf{Precision} & \textbf{Recall}\\
         
    \hline
    \multicolumn{6}{c}{\textbf{DNN}}\\
    \hline
    \textbf{wideband} & 0.70 & 0.45 & 0.75 & 0.70 & 0.70\\
    \textbf{Narrowband} & 0.72 & 0.46 & 0.78 & 0.72 & 0.72\\
    \textbf{Wavelet} & 0.61 & 0.42 & 0.73 & 0.61 & 0.61\\
    \textbf{Early-Fusion (x2)} & 0.65 & 0.38 & 0.73 & 0.66 & 0.65 \\
    \textbf{Late-Fusion (x2)} & 0.66 & 0.39 & 0.71 & 0.68 & 0.66 \\
    \textbf{Multi-spectral Fusion (x2)} & \textit{0.92} & \textit{0.70} & \textit{0.98} & \textit{0.92} & \textit{0.92} \\
    
    \hline
    \multicolumn{6}{c}{\textbf{SVM}}\\
    \hline
    \textbf{wideband} & 0.71 & 0.49 & 0.76 & 0.71 & 0.71 \\
    \textbf{Narrowband} & 0.71 & 0.49 & 0.77 & 0.71 & 0.71 \\
    \textbf{Wavelet} & 0.69 & 0.50 & 0.78 & 0.69 & 0.69\\
    \textbf{Early-Fusion (x2)} & 0.70 & 0.51 & 0.77 & 0.70 & 0.70\\
    \textbf{Early-Fusion (x3)} & 0.70 & 0.52 & 0.76 & 0.70 & 0.70\\
    \textbf{Late-Fusion (x2)} & 0.72 & 0.50 & 0.76 & 0.72 & 0.72\\
    \textbf{Late-Fusion (x3)} & 0.70 & 0.51 & 0.76 & 0.70 & 0.70\\
    \textbf{Multi-spectral Fusion (x2)} & \textbf{\textit{0.95}} & \textbf{\textit{0.75}} & \textbf{\textit{0.99}} & \textbf{\textit{0.95}} & \textbf{\textit{0.95}}\\
    \textbf{Multi-spectral Fusion (x3)} & 0.83 & 0.64 & 0.88 & 0.83 & 0.83\\
    \hline
   
    \end{tabular}}
    \label{pd_results_table1}
    
\end{table}

Note that the representations based solely on the wideband or narrowband spectrograms yield similar accuracies. However, the accuracy highly improves when the fusion based on the multi-spectral autoencoder is considered. This fact is reflected with both the DNN and the SVM classifiers, where accuracies up to 95\% are obtained classifying PD and HC subjects. These results outperformed those observed in literature when the same data is considered~\cite{JCVC_parallelRepresentationLearning,godino2017,moro2019forced,karan2019parkinson,lopez2019assessing,rueda2019}. The multi-spectral fusion also produces the best result evaluating the dysarthria severity of the patients ($\rho=0.75$). These results are explained since while both the wideband and narrowband representations individually have a higher resolution in either frequency or time, each representation sacrifices valuable speech-related information due to the lack of resolution in the alternative, respective domain. Moreover, note that for the DDK exercises, both representations are vital. For instance, the wideband signal with higher time resolution, is better at recognizing stop consonants, such as the 'p', 't', or 'k' but is not as accurate representing vowel sounds. On the other hand, the narrowband representation has a higher frequency resolution, which is better to represent vocalic sounds. This is also likely another contributing factor to the high performance of the multi-spectral autoencoder. 

Regarding the wavelet representation, the results are slightly lower than the observed for the wideband and the narrowband spectrograms. No real discernible advantage was found when including it in the early, late, or multi-spectral fusion approaches. The lack of benefit could be due to the nature of the DDK task, where the utterances are more or less consistent and the benefit of the wavelet transform (i.e., adaptability to a signal where resolution needs to be sensitive to variability) is mitigated.

\begin{figure}[ht!]
    \centering
    \includegraphics[width=\linewidth]{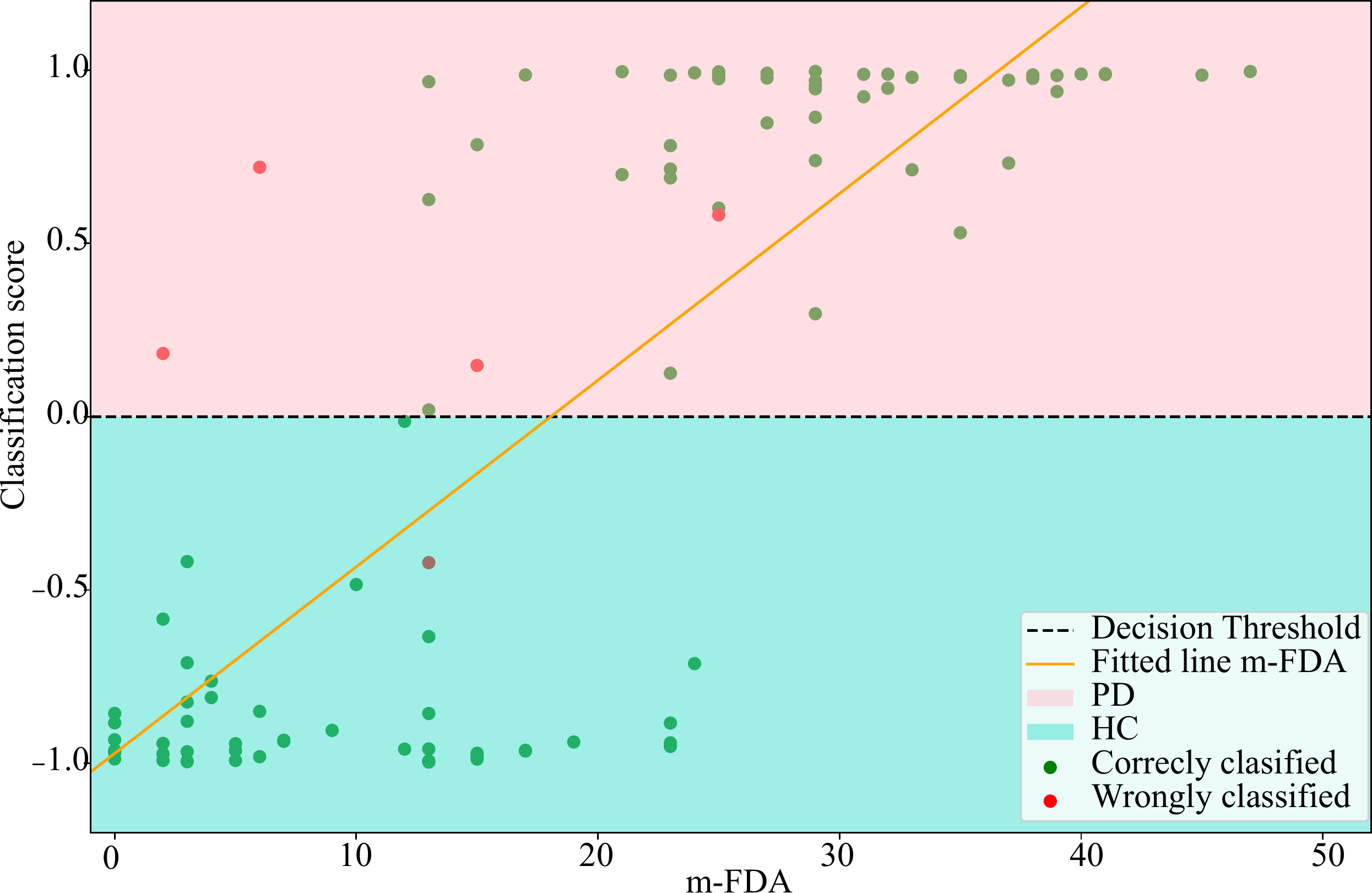}
    \caption{Decision score of the SVM classifier with the multi-spectral autoencoder features against the m-FDA scores of the participants. Green and red dots indicate samples that were correctly and wrongly classified, respectively.}
    \label{svm_mcf_class_res}
\end{figure}

Figure~\ref{svm_mcf_class_res} shows the decision boundary of the SVM classifier when considering the features extracted in the multi-spectral autoencoder. It produced the best result both detecting the presence of the disease and evaluating the dysarthria severity in the participants. The horizontal axis corresponds to the m-FDA score of the subjects, while the vertical axis is the scaled distance to the separating hyperplane according to the Platt scaling~\cite{platt1999probabilistic}. The figure indicates the regions of the samples to be predicted as PD (pink) and HC (cyan), and whether the prediction was accurate (green dots) or not (red dots).  The Spearman correlation ($\rho=0.75$) is computed between the m-FDA scores and the classification score.

To evaluate the generalization capability of the multi-spectral autoencoder, an SVM classifier was trained on the full PC-GITA dataset. Then, using the features extracted with the multi-spectral autoencoder, a total hidden and blind test was conducted on the PD independent dataset, similar to the experiment performed in~\cite{karan2020hilbert}. The results obtained in this experiment are shown in Table~\ref{pd_results_table2}.

\begin{table}[ht!]

    \caption{Classification results for PD patients in the PD independent dataset using the multi-spectral autoencoder combining the wideband and narrowband spectrograms.}

    \centering
    \scriptsize
    \resizebox{\linewidth}{!}{
    \begin{tabular}{lccc}
    \hline
         & \textbf{Accuracy}  &  \textbf{Precision} & \textbf{Recall}\\
    \hline
    \textbf{Multi-spectral Fusion (x2)} & 0.78 & 0.78 & 0.78\\
    \hline
   
    \end{tabular}}
    \label{pd_results_table2}
    
\end{table}

The results are lower than the ones obtained in the cross validation tests with the PC-GITA corpus. This is expected because of the difference in the acoustic conditions among the training and test set. The results obtained here are similar to the reported ones in~\cite{karan2020hilbert}, and are in fact slightly higher depending on the speech task considered for the analysis.

\subsection{Cleft Lip and Palate}

Table~\ref{clp_results_table} shows the results obtained discriminating between the CLP patients and their HC subjects with the autoencoder features extracted from the speech representations using the SVM classifier. Interestingly, the narrowband representation yielded the best results (93\% accuracy, 0.96 AUC, 0.92 precision, 0.93 recall) for this particular problem. The multi-spectral approach performed similarly to the other fusion-based methods and is only slightly less accurate than when using only the narrowband spectrogram.

\begin{table}[ht!]
\caption{Classification results for CLP patients. Early and late fusion results for the SVM classifier are reported for the combination of wideband and narrowband features (x2) and (x3) implies the inclusion of the wavelet-based features.}
    \centering
    \scriptsize
    \begin{tabular}[]{lccccc}
    \hline
         & \textbf{Accuracy}  & \textbf{AUC} & \textbf{*Precision} & \textbf{*Recall}\\
         
    \hline
    \textbf{SVM} & & & &\\
    \hline
    \textbf{wideband} & 0.86 & 0.93 & 0.84 & 0.87\\
    \textbf{Narrowband} & \textbf{\textit{0.93}} & \textbf{\textit{0.96}} & \textbf{\textit{0.92}} & \textbf{\textit{0.93}} \\
    \textbf{Wavelet} & 0.80 & 0.87 & 0.78 & 0.80\\
    \textbf{Early-Fusion (x2)} & 0.89 & 0.94 & 0.87 & 0.89\\
    \textbf{Early-Fusion (x3)} & 0.88 & 0.93 & 0.86 & 0.89 \\
    \textbf{Late-Fusion (x2)} & 0.89 & 0.94 & 0.88 & 0.88 \\
    \textbf{Late-Fusion (x3)} & 0.89 & 
    
    \textbf{\textit{0.96}} & 0.88 & 0.88  \\
    \textbf{Multi-spectral Fusion (x2)} & 0.89 & 0.93 & 0.88 & 0.88\\
    
    \hline
   
    \end{tabular}
    \label{clp_results_table}
\end{table}


When comparing the wideband and narrowband approaches, it is important to recall the speech-related characteristics each exploits. With the wideband approach, the higher time resolution allows for better analysis and tracking of the formants. However, the AE assigns more weight to the reconstruction of lower frequencies, where the majority of energy related to speech lies~\cite{JCVC_parallelRepresentationLearning}. The results indicate that perhaps during the reconstruction of the wideband representation, the representation loses some information in high frequencies, like high order formants which are important classifying CLP patients. On the other hand, the narrowband representation is more robust in representing certain aspects, such as the fundamental frequency, which is also important to detect excessive nasalization. Specifically, it has been shown that that hypernasal speech was rated more monotonous as hypernasality increased, and perception of intonation is lower. Thus, when differentiating speech of CLP speakers and HC speakers, the narrowband signal, and its better characterization of intonation, better discriminates the two cohorts \cite{MTB_hypernasalSpeechMoreMonotonous}.

\section{Conclusion}

In this paper, the topic of automated disease prognosis and classification based on the speech of PD and CLP patients was explored. Two main aspects of a given disease or disorder were specifically considered: the presence of a disease/disorder via classification of HC subjects and patients as well as the level of degradation of the speech of patients according to a specific clinical scale (in the case of PD speakers). An unsupervised method based on feature representation learning was employed to extract robust features, which helped to improve the accuracy of different models to classify pathological speech. The strategy was adapted from an autoencoder based framework~\cite{JCVC_parallelRepresentationLearning}. This study focused on variations of the signal input to that model. Specifically, three different speech representation types were considered, the wideband and narrowband Mel-spectrograms and wavelet scalogram. In addition, fusion-based feature aggregation methods, which combines information from each representation, were also considered. One such aggregation approach, the multi-spectral autoencoder was introduced for the first time and shown to be highly capable in identifying PD and assessing the dysarthria severity of the participants.

For the case of evaluating PD patients, the multi-spectral autoencoder yielded the best results. Further analysis was done to evaluate the classification method on an independent data set. Results were again promising. The accuracy achieved of 78\% is higher than that observed in a similar study~\cite{BK_hilbertSpectrumAnalysisAutoDetectionPD}. These results are in fact indicative that the proposed model may be suitable to perform automatic detection of PD speakers in real-world conditions. On the other hand, for CLP patients, the narrowband representation yielded the best results (accuracy of 93\%). The results here were competitive with similar approaches from literature~\cite{JCVC_parallelRepresentationLearning}.

Looking forward, much can still be done to further analyze the approaches proposed in this work with the goal of reducing the subjectivity of the clinical evaluation process for PD and CLP patients. When considering the specific representations, different parameterizations could also be considered, i.e., increasing or decreasing the analysis window and the number of frequency bands in order to optimize the amount of information obtained by either the wideband or narrowband representations. In regards to the work done with CLP patients, this research did not specifically consider a severity assessment of the patients, something that could be done in future work. Beyond PD and CLP speakers, the proposed approach could also be extended to test its ability to evaluate patients with other speech disorders, such as cochlear implant users, or those stemming from neck cancer, or depression for example. It could also be of interest to compare a given classifier's ability to discern different neurodegenerative diseases.

\section{Acknowledgements}

This project received funding from the EU Horizon 2020 research and innovation programme under the Marie Sklodowska-Curie Action ITN-ETN project TAPAS, Grant Agreement No. 766287. 
This work was partially funded by CODI from the University of Antioquia, grant \# PRG2017-15530.

\bibliographystyle{ieeetr}
\bibliography{main}
\end{document}